\documentclass[a4paper]{scrartcl}

\usepackage[utf8]{inputenc}
\usepackage[english]{babel}
\usepackage[T1]{fontenc}
\usepackage{listings}
\usepackage{csquotes}

\usepackage{mathtools}
\usepackage{bm,dsfont}
\usepackage{ntheorem}

\usepackage{array,booktabs}

\usepackage{listings}
\usepackage[usenames,dvipsnames]{xcolor}

\lstdefinelanguage{Julia}%
  {morekeywords={abstract,break,case,catch,const,continue,do,else,elseif,%
      end,export,false,for,function,immutable,import,importall,if,in,%
      macro,module,otherwise,quote,return,switch,true,try,type,typealias,%
      using,while,julia},%
   sensitive=true,%
   alsoother={$},%
   morecomment=[l]\#,%
   morecomment=[n]{\#=}{=\#},%
   morestring=[s]{"}{"},%
   morestring=[m]{'}{'},%
}[keywords,comments,strings]%

\usepackage{graphicx}
	\graphicspath{{figures/}}
\usepackage{todonotes}

\usepackage{xcolor}
\definecolor{aaublue}{rgb}{0.00,0.41,0.66}
\definecolor{nicered}{rgb}{.647,.129,.149}

\usepackage{hyperref}
\hypersetup{%
	breaklinks=true,%
	colorlinks=true,%
	linkcolor=aaublue,%
	citecolor=nicered,%
	urlcolor=aaublue,%
	pdftitle={Fast Generalized Sampling},%
	pdfauthor={Robert Dahl Jacobsen et al},%
	pdfsubject={Generalized sampling},%
	pdfkeywords={Generalized sampling},%
	pdfpagemode=UseOutlines,%
	pdfstartview=FitH%
}

\newcommand{\vv}[1]{\ensuremath{\bm{#1}}}

\newcommand{\F}{\ensuremath{\mathcal{F}}} 

\newcommand{\Hil}{\ensuremath{\mathcal{H}}} 

\DeclarePairedDelimiterX\inner[2]{\langle}{\rangle}{ #1 , #2 }

\DeclarePairedDelimiter\norm{\lVert}{\rVert}

\DeclarePairedDelimiter\abs{\lvert}{\rvert}

\newcommand{\ind}[1]{\ensuremath{\mathds{1}_{#1}}}

\newcommand{\Space}{\ensuremath{\quad}}
\newcommand{\FOR}[1]{\ensuremath{\Space\text{#1}}}

\newcommand\samp{\ensuremath{s}}

\newcommand\recon{\ensuremath{r}}

\newcommand{\app}[1]{\ensuremath{\widetilde{#1}}}

\newcommand{\A}{\ensuremath{*}}

\newcommand{\w}{\ensuremath{w}}

\newcommand\cb{\ensuremath{T}}
\newcommand\cbe{\ensuremath{t}}

\DeclareMathOperator\diag{diag}

\DeclareMathOperator\NDFT{NDFT}
\DeclareMathOperator\NFFT{NFFT}

\DeclareMathOperator{\SPAN}{span}

\newcommand\pinv{\ensuremath{\dagger}}

\DeclareMathOperator*{\argmin}{argmin}

\DeclareMathOperator{\supp}{supp}

\newcommand{\had}{\circ}

\newcommand\given{\:\vert\:}
\newcommand\SetSymbol[1][]{
	\nonscript\: #1 \vert \nonscript\:
	\mathopen{} 
	\allowbreak 
}
\DeclarePairedDelimiterX\Set[1]{\{}{\}}{
	\renewcommand\given{\SetSymbol[\delimsize]}
	#1
}

\newcommand{\N}{\ensuremath{\mathds{N}}}
\newcommand{\Z}{\ensuremath{\mathds{Z}}}
\newcommand{\R}{\ensuremath{\mathds{R}}}
\newcommand{\C}{\ensuremath{\mathds{C}}}

\newcommand{\lphi}{\ensuremath{\phi^{\text{L}}}}
\newcommand{\rphi}{\ensuremath{\phi^{\text{R}}}}
\newcommand{\iphi}{\ensuremath{\phi^{\text{int}}}}

\newcommand\lH{\ensuremath{H^{\text{left}}}}
\newcommand\lh{\ensuremath{h^{\text{left}}}}
\newcommand\rH{\ensuremath{H^{\text{right}}}}
\newcommand\rh{\ensuremath{h^{\text{right}}}}

\newcommand\lU{\ensuremath{U_{\text{left}}}}
\newcommand\lV{\ensuremath{V_{\text{left}}}}
\newcommand\lv{\ensuremath{\vv v^{\text{left}}}}

\newcommand\rU{\ensuremath{U_{\text{right}}}}
\newcommand\rV{\ensuremath{V_{\text{right}}}}
\newcommand\rv{\ensuremath{\vv v^{\text{right}}}}

\newcommand{\Left}{L}
\newcommand{\Right}{R}
\newcommand{\Int}{I}

\DeclareMathOperator\vecto{vec}
\DeclareMathOperator\idx{idx}

\theoremheaderfont{\normalfont\bfseries}
\theorembodyfont{\normalfont}
\theoremstyle{break}

\newtheorem{theorem}{Theorem}

\title{Generalized Sampling in Julia}
\author{Robert Dahl Jacobsen\thanks{Department of Mathematical Sciences, Aalborg University, Fredrik Bajers Vej 7G, DK-9220 Aalborg East} \and Morten Nielsen\footnotemark[\value{footnote}]{} \and Morten Grud Rasmussen\footnotemark[\value{footnote}]}

\begin{document}
\maketitle

\footnotetext{Supported by the Danish Council for Independent Research | Natural Sciences, grant 12-124675, "Mathematical and Statistical Analysis of Spatial Data".
RDJ is supported by the Centre for Stochastic Geometry and Advanced Bioimaging, funded by a grant (8721) from the Villum Foundation.
}

\begin{abstract}
Generalized sampling is a numerically stable framework for obtaining reconstructions of signals in different bases and frames from their samples.

In this paper, we  will introduce a carefully documented  toolbox for performing generalized sampling in Julia. Julia is a new language for technical computing with focus on performance, which is ideally suited to handle  the large size problems often encountered in generalized sampling. The toolbox provides specialized solutions for the setup of Fourier bases and wavelets.

The performance of the toolbox is compared to existing implementations of generalized sampling in MATLAB.
\end{abstract}


\section{Introduction}

Generalized sampling \cite{Adcock:Hansen:2012:GS,Adcock:Hansen:Poon:2014} is a framework for estimating representations of functions in different bases and frames in a numerically stable manner.
This paper documents a toolbox for performing generalized sampling in Julia \cite{Bezanson:Edelman:Karpinski:Shah:2014}.
Julia is a new language for technical computing with focus on performance, which is essential for the large problems encountered in generalized sampling.

The theory of generalized sampling does not restrict the type of bases to consider, but the applications have focused on Fourier bases and multiscale representations like wavelets.
Hence our software has specialized solutions for this particular set of bases. 

The paper is organized as follows: In
Section\nobreakspace \ref {sec:generalized_sampling_intro} and Section\nobreakspace \ref {sec:prereq} we recap
the generalized sampling framework, set the notation and discuss the
prerequisites.  In Sections\nobreakspace \ref {sec:sampling1D} and\nobreakspace  \ref {sec:sampling2D} we introduce
the algorithms in 1D and 2D, respectively.
Section\nobreakspace \ref {sec:GS_package} is an introduction to the released software.

\section{Generalized sampling}
\label{sec:generalized_sampling_intro}

Mathematically, samples of a function $f$ in a Hilbert space $\Hil$ with respect to a sample basis $\Set{\samp_n}_{n\in\N}$ consists of inner products $\Set{\inner f{\samp_n}}_{n\in\N}$.
In generalized sampling we want to use these samples to estimate the inner products $\Set{\inner f{\recon_n}}_{n\in\N}$, where $\Set{\recon_n}_{n\in\N}$ is another basis for $\Hil$.
The basis $\Set{\recon_n}_{n\in\N}$ is used for reconstructing $f$.

In practice we only consider a finite number of sampling and reconstruction functions, i.e., we have access to the samples
\begin{equation*}
	\w^\samp_m = \inner f{\samp_m}, \FOR{$1\leq m\leq M$}.
\end{equation*}
From these samples we estimate the coefficients in the reconstruction basis,
\begin{equation}
	\label{eq:GS_reconstruction_coefficients}
	\app\w^\recon_n \approx \w^\recon_n = \inner f{\recon_n}, \FOR{$1\leq n\leq N$}
\end{equation}
which are used to compute an approximation of $f$,
\begin{equation}
	\label{eq:GS_reconstruction}
	\app f_{N,M} = \sum_{n=1}^N \app\w^\recon_n \recon_n.
\end{equation}

The actual computation of the reconstruction coefficients $\app{\vv\w}^r = \Set{\app\w^\recon_n}_{n=1}^N$ is performed by solving a least squares problem.
The infinite change of basis matrix between the sampling and reconstruction subspaces has $(i,j)$'th entry $\inner{\recon_j}{\samp_i}$.
We consider a finite $M\times N$ section of this matrix, denoted by \cb:
\begin{equation}
	\label{eq:change_of_basis_matrix}
	\cb = \bigl[\inner{\recon_j}{\samp_i}\bigr]_{1\leq j\leq N}^{1\leq i\leq M}.
\end{equation}
The reconstruction coefficients are computed as the least squares solution
\begin{equation}
	\label{eq:least_squares}
	\app{\vv\w}^r = \argmin\Set[\big]{\norm{\cb \vv x - \vv\w^s}_2 \given \vv x\in \C^N}.
\end{equation}
It is well-known that the solution of \textup {(\ref {eq:least_squares})} is $\app{\vv\w}^r = \cb^\pinv \vv\w^s$ where $\cb^\pinv$ is the pseudo-inverse of $\cb$.
However, for large matrices \cb\ it is not feasible to compute this solution analytically.
In fact, for realistic sample sizes it may not even be possible to store the change of basis matrix \textup {(\ref {eq:change_of_basis_matrix})}.
Therefore, in order to enjoy generalized sampling we need specialized algorithms for each set of sampling and reconstruction bases.

A popular algorithm for solving large least squares problems is conjugate gradients (see e.g.\ \cite[p. 637]{Golub:van_Loan:2013}).
To apply the conjugate gradients algorithm we need to be able to compute matrix-vector products with \cb\ and its adjoint $\cb^*$.
The convergence rate of conjugate gradients (and similar algorihtms) depends on the condition number of \cb. 
As mentioned earlier, the benefit of generalized sampling is that we have conditions that ensure good numerical properties of $\cb$.
The key trick that ensures numerical stability is to let $N < M$, i.e., to have more samples than reconstruction coefficients; 
the relation between $N$ and $M$ is determined by the \emph{stable sampling rate}.
For the particular choice of sampling in Fourier space and reconstructing with wavelets the stable sampling rate is linear \cite{Adcock:Hansen:Poon:2014}.

Thus, in order to perform generalized sampling, i.e., compute the desired coefficients \textup {(\ref {eq:GS_reconstruction_coefficients})} we need to be able to perform multiplications with $\cb$ and $\cb^*$.
For the specific choice of sampling with Fourier bases and reconstructing in wavelet bases, this can be accomplished efficiently by using non-uniform fast Fourier transforms.

\section{Prerequisites}
\label{sec:prereq}
\subsection{Notation}
\label{sec:notation}

When sampling frequency responses we let $\Set{\xi_m}_{m=1}^M$ denote the frequency locations in 1D and $\Set{\vv\xi_m}_{m=1}^M$, where $\vv\xi_m = (\xi_{x_m}, \xi_{y_m})$, denote the frequency locations in 2D.
With $\Hil = L^2(\R^d)$ the elements in \cb\ are evaluations of the Fourier transforms of the reconstruction functions, i.e., $\inner{\recon_j}{\samp_i} = \widehat\recon_j(\xi_i)$.

\subsection{Non-uniform fast Fourier transform}
\label{sec:NFFT}

To introduce the non-uniform discrete Fourier transform (NDFT) in dimension $D$, let $N_d$, $d=1,\ldots,D$ be even, positive integers, $\vv N = (N_1,\ldots,N_D)$ and 
\begin{equation*}
	I_{\vv N} = \Z^D \cap \prod_{d=1}^D \Bigl[-\frac{N_d}2, \frac{N_d}2\Bigr),
\end{equation*}
The set of sampling locations is denoted $\chi = \Set{\vv\xi_m}_{m=1}^M$, where $\vv\xi_m \in [-\frac12, \frac12)^D$ and we let $M$ denote the size of $\chi$.
The NDFT of $\vv x = \Set{x_{\vv k} \in \C \given \vv k \in I_{\vv N}}$ with sampling locations $\chi$ is denoted $\vv y = \NDFT[\chi](\vv x)$ where
\begin{equation}
	\label{eq:NDFT_def}
	y_m
	= \sum_{\vv k\in I_{\vv N}} x_{\vv k} \exp\bigl( -2\pi i \vv k \cdot \vv\xi_m \bigr),
	\FOR{$m = 1,\ldots,M$}.
\end{equation}
This can be written as a matrix multiplication $\vv y = F\vv x$, where $F_{m,n} = \exp(-2\pi i \vv k_n\cdot \vv\xi_m)$ with a suitable ordering of the elements in $I_{\vv N}$.

The adjoint NDFT is defined as multiplication with $F^\A$, i.e., if $\vv z = F^\A\vv y$, then
\begin{equation*}
	z_{\vv k}
	= \sum_{m=1}^M y_m \exp\bigl( 2\pi i \vv k \cdot \vv\xi_m \bigr),
	\FOR{$\vv k \in I_{\vv N}$}.
\end{equation*}

In Julia we have access to an NFFT package (\url{https://github.com/tknopp/NFFT.jl}) for fast \emph{approximation} of the NDFT and adjoint NDFT.
The NFFT package is inspired by the C library documented in \cite{Keiner:Kunis:Potts:2009}.

If the $M$ sampling points are uniformly distributed, the NDFT \textup {(\ref {eq:NDFT_def})} reduces to a DFT with appropriate phase shifts.
For simplicity we consider the 1D situation.
In the setup that is relevant for our generalized sampling application the sampling locations are
\begin{equation*}
	\xi_m 
    = \frac\varepsilon{N} \biggl(m - 1 - \frac M2\biggr),
\end{equation*}
and the signal to be transformed is $x_1, \cdots, x_{N_1}$, where $M \geq N \geq N_1$ and $\varepsilon^{-1} \in \N$.
With $\ell = k + 1 + N_1/2$ \textup {(\ref {eq:NDFT_def})} becomes
\begin{align*}
	y_m 
	& = \sum_{k=-N_1/2}^{N_1/2-1} x_{k+1+N_1/2} \exp\bigl(-2\pi i k \xi_m\bigr)
	\\
    & = \exp(\pi i N_1 \xi_m) \sum_{\ell=1}^{N_1} x_\ell \exp\bigl(-2\pi i (\ell-1) \xi_m\bigr)
	\\
    & = \exp(\pi i N_1 \xi_m) \sum_{\ell=1}^{N_1} x_\ell \exp\Bigl(\pi i (\ell-1) \frac{\varepsilon M}N\Bigr) \exp\biggl(-2\pi i \frac{(\ell-1) (m-1)}{N/\varepsilon}\biggr).
\end{align*}
Let $q = \min\Set{p \in \N \given p N \varepsilon^{-1} \geq M}$ and $N_2 = q N \varepsilon^{-1}$.
Furthermore, let $\vv z$ be a vector of length $N_2$ where
\begin{equation*}
	z_n = 
	\begin{dcases}
		x_\ell \exp\Bigl(\pi i (\ell-1) \frac{\varepsilon M}N\Bigr), & n = q(\ell - 1), \Space 1 \leq \ell\leq N_1,
		\\
        0, & \text{otherwise}.
	\end{dcases}
\end{equation*}
With this notation we see that
\begin{equation*}
	y_m 
    = \exp(\pi i N_1 \xi_m) \sum_{n=1}^{N_2} z_n \exp\biggl(-2\pi i \frac{(n-1) (m-1)}{N_2}\biggr).
\end{equation*}
This sum is one entry in the DFT of $\vv z$.
So $\NDFT[\chi](\vv x)$ can be computed from the first $M$ entries of the DFT of $\vv z$.

\subsection{Daubechies scaling functions}
\label{sec:Daubechies_scaling_functions}

Let $\phi$ denote the scaling function of a multiresolution analysis (see e.g.\ \cite{Hernandez:Weiss:1996}).
The scaling function on scale $J$ with translation $k$ is defined as
\begin{equation*}
	\phi_{J,k}(x) 
	= 2^{J/2} \phi(2^J x - k).
\end{equation*}
We know that
\begin{equation*}
	L^2(\R) 
	= \overline{ \bigcup_{J\in\Z} V_J },
	\FOR{where }
	V_J = \SPAN\Set{\phi_{J,k} \given k\in \Z}
\end{equation*}
If $\phi$ is the Haar wavelet, then for all $J_0 \geq 0$
\begin{equation*}
	\overline{ \bigcup_{J\geq J_0} \SPAN\Set{\phi_{J,k} \given -2^{J-1}\leq k< 2^{J-1}} }
	= L^2([-\tfrac12, \tfrac12]).
\end{equation*}
For Daubechies wavelets of higher orders, $J$ needs to be large enough to ensure that $\supp(\phi_J) \subseteq [-\frac12, \frac12]$ and the functions near the boundaries (i.e., with $k$ near $-2^{J-1}$ or $2^{J-1}$) need further corrections.
We use the boundary wavelets of \cite{Cohen:Daubechies:Vial:1993} that have the same number of vanishing moments as the internal/non-boundary wavelets.
With $p$ vanishing moments there are $p$ left boundary functions $\lphi_k$ and $p$ right boundary functions $\rphi_k$, $k=0,\ldots,p-1$.
In both cases $k=0$ is the function closest to the associated edge, i.e., when traversing the functions from left to right the order is $\lphi_0, \ldots, \lphi_{p-1}$ at the left edge and $\rphi_{p-1}, \ldots, \rphi_0$ at the right edge.
At scale $J$ we define
\begin{equation*}
	\lphi_{J,k}(x) 
	= 2^{J/2} \lphi_k(2^J x)
\end{equation*}
and similarly for $\rphi_k$.

Let $\tau_h$ deonte the translation operator, $(\tau_h f)(x) = f(x - h)$.
For a scaling function related to a Daubechies wavelet with $p>1$ vanishing moments, $\supp(\phi) = [-p+1, p]$ and for $J$ with $2^J \geq p$, we let
\begin{equation*}
	\iphi_{J,k}(x) =
	\begin{dcases}
		\bigl(\tau_{-\frac12}\lphi_{J,2^{J-1}+k}\bigr)(x), & -2^{J-1}\leq k< -2^{J-1}+p,
		\\
		\phi_{J,k}(x), & -2^{J-1}+p\leq k < 2^{J-1}-p,
		\\
		\bigl(\tau_{\frac12}\rphi_{J,2^{J-1}-1-k}\bigr)(x), & 2^{J-1}-p\leq k < 2^{J-1}.
	\end{dcases}
\end{equation*}
For the Haar wavelet $\iphi_{J,k} = \phi_{J,k}$.
Let now 
\begin{equation}
	\label{eq:modified_Daubechies_span}
	V_J^{\text{int}}
	= \SPAN\Set[\big]{ \iphi_{J,k} \given -2^{J-1}\leq k< 2^{J-1} }.
\end{equation}
Then 
\begin{equation*}
	L^2([-\tfrac 12,\tfrac 12])
	= \overline{ \bigcup_{J\geq\log_2(2p)} V_J^{\text{int}} }.
\end{equation*}

\section{Generalized sampling in 1D}
\label{sec:sampling1D}

For a function $f \in L^2(\R)$ we wish to compute an approximation of $\smash{f\ind{[-\frac12, \frac12]}}$ with scaling functions from a single $V_J^{\text{int}}$ as defined in \textup {(\ref {eq:modified_Daubechies_span})}.
Let $\chi = \{\xi_m\}_{m=1}^M$ denote the frequency locations where we obtain samples $y_m = f(\xi_m)$.
The scale of the reconstruction is $J$ and $N = 2^J$ is the number of reconstructed coefficients.
Let $p \geq 1$ be the number of vanishing moments of the scaling function and $\cbe_{m,n}$ denote the $(m,n)$'th entry of the change of basis matrix \cb:
\begin{equation*}
	\cbe_{m,n} 
	= \F\bigl[\iphi_{J,n-1-2^J}\bigr](\xi_m) = 
	\begin{dcases}
		\F[\tau_{-\frac12}\lphi_{J,n-1}](\xi_m), & 1\leq n\leq p,
		\\
		\F[\phi_{J,n-1-2^{J-1}}](\xi_m), & p< n \leq 2^J-p,
		\\
    \F[\tau_{\frac12} \rphi_{J,2^J-n}](\xi_m), & 2^J-p< n\leq 2^J.
	\end{dcases}
\end{equation*}
With the usual calculus for Fourier transforms, \textup {(\ref {eq:Fourier_dilation})} and\nobreakspace  \textup {(\ref {eq:Fourier_translation})}, we have that
\begin{align*}
	\F[\tau_{-1}\lphi_{J,n-1}](\xi_m)
	& = 2^{-J/2} \exp\bigl(+\pi i \xi_m\bigr) \F[\lphi_{n-1}] \bigl(2^{-J} \xi_m\bigr),
	\\
	\F[\phi_{J,n-1-2^{J-1}}](\xi_m)
	& = 2^{-J/2} \exp\bigl(-\pi i (n-1-2^{J-1}) 2^{-J} \xi_m\bigr) \F[\phi]\bigl(2^{-J} \xi_m\bigr),
	\\
	\F[\tau_1 \rphi_{J,n-1}](\xi_m)
	& = 2^{-J/2} \exp\bigl(-\pi i \xi_m\bigr) \F[\rphi_{2^J-n}] \bigl(2^{-J} \xi_m\bigr).
\end{align*}
Introduce the diagonal matrix
\begin{equation*}
	D 
	= \diag\Bigl( 2^{-J/2} \widehat\phi\bigl(2^{-J} \xi_m\bigr), 1\leq m\leq M\Bigr)
\end{equation*}
and the NDFT matrix $F$ of size $M\times N$ with $(m,n)$'th entry
\begin{equation*}
	F_{m,n} 
	= \exp\bigl(-2\pi i (n-1-2^J) 2^{-J} \xi_m\bigr)
\end{equation*}
Since $2^{J-1} = N/2$ the multiplication $DF\vv x$ can be approximated as $D\cdot \NFFT[\chi](\vv x)$.
Let $\Left$ be the $M\times p$ matrix with entries $L_{m,n} = \cbe_{m,n}$, $\Right$ be the $M\times p$ matrix with entries $R_{m,n} = \cbe_{m,N-p+1+n}$ and $\Int = D\cdot F$.
With this notation we can write \cb\ as the block matrix
\begin{equation}
	\label{eq:CoB_block_1D}
	\cb = 
	\begin{bmatrix}
		\Left & \Int & \Right
	\end{bmatrix}.
\end{equation}

\subsection{Non-uniform sampling}
\label{sec:non-uniform_sampling}

With non-uniform sampling points we may have clusters and desolate areas in the frequencies.
To compensate for this the reconstructed coefficients are computed as the solution of a \emph{weighted} least squares problem:
\begin{equation}
	\label{eq:weighted_least_squares}
	\app{\vv\w}^r 
	= \argmin\Set[\big]{\norm{Q(\cb \vv x - \vv\w^s)}_2 \given \vv x\in \C^N},
\end{equation}
where $Q = \diag(\mu_m, 1\leq m\leq M)$.
From the solver's point of view, the only change from the ordinary least squares is that $\w^s_m$ and $\cbe_{m,n}$ are multiplied with $\sqrt{\mu_m}$ for all $n=1,\ldots,N$.

To determine the weights, we follow \cite{Adcock:Gataric:Hansen:2014}
and \cite{Adcock:Gataric:Hansen:2015}. For simplicity, we focus on the
case where the unknown function $f$ is supported in $[-\frac12, \frac12]$, $f\in
L^2([-\frac12, \frac12])$. Define the
(inverse\footnote{\cite{Adcock:Gataric:Hansen:2015} refers to is as
a \emph{density}, but the lower the number, the more dense is the
set}) $([-\frac12, \frac12], \Omega, Y)$-\emph{density} $\delta_{[-\frac12, \frac12]}(\Omega,
Y)=\frac{1}{2}\sup_{y\in Y}\inf_{\xi\in\Omega}\abs{\xi-y}$, where $\Omega$ is the
set of sampling frequencies and $Y\subset\R$ is a closed, simply
connected set.

With this notation, we sum up some of the results
of \cite{Adcock:Gataric:Hansen:2015}:
\begin{theorem}
   Let $\Omega$ be a countable set of sampling frequencies such that
   $\smash{\delta_{[-\frac12,
   \frac12]}}(\Omega, \R)<\frac{1}{4}$. Then $\{\sqrt{\mu_\omega}e_\omega\}_{\omega\in\Omega}$
   is a weighted Fourier frame for $L^2([-\frac12, \frac12])$, where
   $e_\omega(x)=e^{i2\pi\omega x} \ind{[-\frac12,\frac12]}(x)$ and
   $\mu_\omega$ is the Lebesque measure of the Voronoi region of
   $\omega\in\Omega$.
\end{theorem}

\begin{theorem}
   Consider $V_J^{\mathrm{int}}\subset L^2([-\frac12, \frac12])$ and let
   $\{\sqrt{\mu_\omega}e_\omega\}_{\omega\in\Omega}$ be a weighted
   Fourier frame with frame bounds $A$ and $B$, where $\Omega$ is a
   countable set of sampling frequencies. Assume that $K$ is closed,
   simply connected set with $0$ in its interior satisfying that
   $\Omega_N=\Omega\cap K$ is finite with cardinality $N$ and
   that \begin{equation*} R(\Omega_N,
   V_J^{\mathrm{int}})=\sup\{\sum_{\xi\in\Omega\setminus\Omega_N}\mu_\xi\abs{\hat
   f(\xi)}^2\,\vert\,f\in
   V_J^{\mathrm{int}}, \norm{f}=1\}<A.  \end{equation*} Then the
   truncated frame operator $S_N$ associated to
   $\{\sqrt{\mu_\omega}e_\omega\}_{\omega\in\Omega_N}$ satisfies that
   $\inner{S_N f}{f}\ge (A-R(\Omega_N,
   V_J^{\mathrm{int}}))\norm{f}^2$, and for every $f\in L^2(\R)$
   there exists a unique $\tilde f=F(f)\in V_J^{\mathrm{int}}$ such
   that \begin{equation*} \forall g\in
   V_J^{\mathrm{int}}\colon\quad\inner{S_Nf}{g}=\inner{S_N\tilde
   f}{g}, \end{equation*} and if $P\colon L^2(\R)\to
   V_J^{\mathrm{int}}$ denotes the orthogonal projection onto
   $V_J^{\mathrm{int}}$, then $F$ satisfies 
   \begin{equation*} 
	   \forall f, h\in
   L^2(\R)\colon\quad\norm{f-F(f+h)}\le\sqrt{\frac{\norm{S_N}}{A-R(\Omega_N,
   V_J^{\mathrm{int}}))}}(\norm{f-Pf}+\norm{h}).
   \end{equation*} 
\end{theorem}

Consequently, if the inverse $([-\frac12, \frac12], \Omega_N, Y)$-density
satisfies $\smash{\delta_{[-\frac12, \frac12]}}(\Omega_N, Y)<\frac{1}{4}$ for some $Y$ and
sufficiently large $N$, the above results leads to \textup {(\ref {eq:weighted_least_squares})}.

\section{Generalized sampling in 2D}
\label{sec:sampling2D}
For $f \in L^2(\R^2)$ we wish to compute an approximation of $\smash{f\ind{[-\frac12, \frac12]\times [-\frac12, \frac12]}}$.
A very natural generalization af the 1D approach to this  2D setting is to use tensor product scaling functions as basis for  $V_J^{\text{int}}\otimes V_J^{\text{int}}$ and obtain an approximation  relative to $V_J^{\text{int}}\otimes V_J^{\text{int}}$.

This way we  obtain a similar block structure of the change of basis matrix. Indeed, if both dimensions are divided as in \textup {(\ref {eq:CoB_block_1D})}, the division of 2D coefficients are
\begin{equation}
	\label{eq:2Dwavelet_border_division}
	\begin{bmatrix}
		\Left_x\Left_y & \Left_x\Int_y & \Left_x\Right_y
		\\
		\Int_x\Left_y & \Int_x\Int_y & \Int_x\Right_y
		\\
		\Right_x\Left_y & \Right_x\Int_y & \Right_x\Right_y
	\end{bmatrix}
\end{equation}
An explicit example of \eqref{eq:2Dwavelet_border_division} in the setup with one sample point and two left, two central, and two right scaling functions can be found in Appendix \ref{2d_mult}. 

Just as in the 1D case, non-uniform sampling patterns require that we solve a weighted least squares problem \textup {(\ref {eq:weighted_least_squares})} and introduce the notion of bandwidth and density -- this deferred to Appendix\nobreakspace \ref {sec:2D_weights_density}.

\section{The GeneralizedSampling package}
\label{sec:GS_package}

This section introduce the basic use of the GeneralizedSampling package through examples and demonstrate the performance.
The examples are shamefully copied from Hansen et al.

\subsection{Package overview}

We have two goals with the GeneralizedSampling package:
It should be fast and easy to use.
We have therefore put effort into providing only a few necessary high-level functions and hiding the lower level details.

The most important function is \lstinline{Freq2Wave} that computes a representation of the change of basis matrix.
Several built-in functions are overloaded to make the output of \lstinline{Freq2Wave} behave like an ordinary matrix, including the backslash operator ``\lstinline{\}'' for computing least squares solutions to $Tx = y$.
Currently, the least squares solution is computed with a conjugate gradient procedure.

A separate package, IntervalWavelets, has been developed to visualize the wavelet representations and is available at \url{https://github.com/robertdj/IntervalWavelets.jl}.
The function of interest from IntervalWavelets is \lstinline{weval} that evaluates a representation in the basis of $V_J^{\text{int}}$ from \textup {(\ref {eq:modified_Daubechies_span})}.

\subsection{Using the package}
\label{sec:GSpackage_usage}

We begin with an example of how reconstruction is performed in 1D.
These examples are also included in the package as scripts that are ready to run.

The Fourier transform $\widehat f$ of function $f: \R \to \R$ is measured in the frequency domain at $\{\frac n2\}_{n=-64}^{63}$, i.e., we have access to $\{\widehat f\bigl(\frac n2\bigr)\}_{n=-64}^{63}$.
For convenience, points on a uniform grid with distance $\varepsilon$ apart are available with the function \lstinline{grid}.
We wish to compute an approximation of $f$ in the Haar basis at scale 6, i.e., with 64 Haar scaling functions.

In Julia, let \lstinline{fhat} denote a vector with the values of the Fourier transform.
\begin{lstlisting}
julia> using GeneralizedSampling
julia> xi = grid(128, 0.5)
julia> T = Freq2Wave(xi, "haar", 6)
julia> wcoef = T \ fhat
\end{lstlisting}

To evaluate the vector \lstinline{wcoef} of coefficients for the Haar scaling functions, \lstinline{weval} of the IntervalWavelets package is used.
The \lstinline{wcoef} vector has complex entries and \lstinline{weval} only accepts real vectors. Furthermore, the resolution of the reconstruction must be specified:
A general Daubechies wavelet can only be computed in the dyadic rationals, i.e., points of the form $k/2^R$ for $k \in \Z$ and $R \in \N \cup \Set{0}$, where $R$ is referred to as the resolution.
\begin{lstlisting}
julia> using IntervalWavelets
julia> x, y = weval(real(wcoef), "haar", 10)
\end{lstlisting}

\begin{figure}
	\centering
	\includegraphics[scale=0.4]{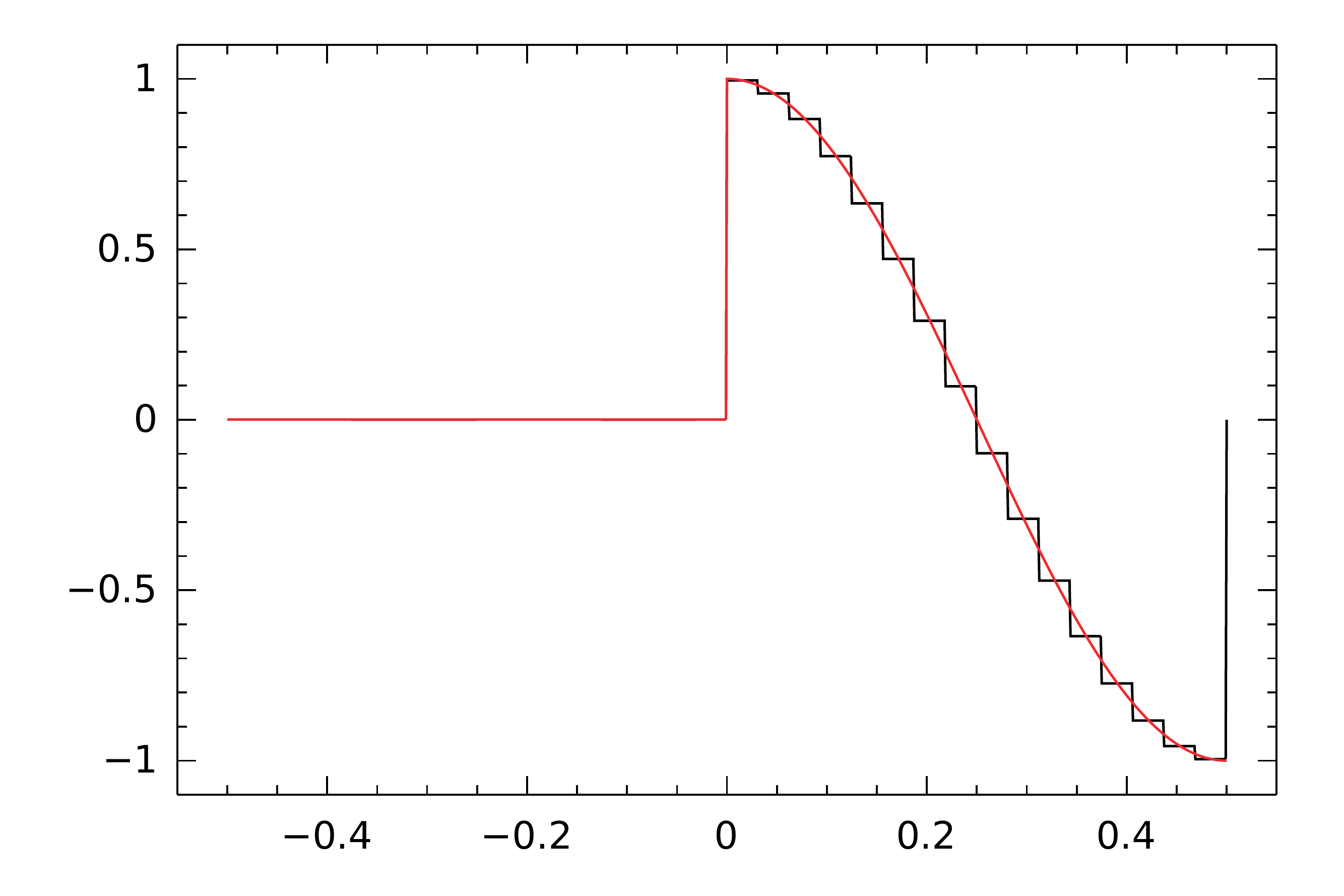}
	\caption{A truncated cosine and approximations with Haar scaling functions.}
	\label{fig:Haar_truncated_cosine}
\end{figure}

An example included in the package is the reconstruction of a truncated cosine (with inspiration from \cite{Gataric:Poon:2016}).
The result is seen in Fig.\nobreakspace \ref {fig:Haar_truncated_cosine}.

To reconstruct in a different Daubechies basis associated with at wavelet with $p$ vanishing moments, two things must be changed in the above code:
\begin{lstlisting}
julia> T = Freq2Wave(xi, "dbp", J)
julia> x, y = weval(real(wcoef), "dbp", 10)
\end{lstlisting}
The output of \lstinline{weval} are vectors with entries that are pairs of $\bigl(x, \app f_{N,M}\bigr(x)\bigr)$.
In the example with the truncated cosine the higher order, continuous Daubechies scaling functions are not well suited to represent the discontinuity.

As metioned in Section\nobreakspace \ref {sec:non-uniform_sampling} it may be of interest to compute reconstructions from non-uniform sampling points.
In this situation the bandwidth must be supplied as a fourth parameter to \lstinline{Freq2Wave}.

Reconstruction of 2D functions/images is performed in a very similar manner.
The only difference is that the sampling locations \lstinline{xi} must be a matrix with two columns.
Remember when choosing the scale $J$ that the number of scaling functions at scale $J$ is $4^J$ instead of the $2^J$ in 1D and the matrices therefore grow rapidly with the scale.

\begin{figure}
	\centering
	\includegraphics[width=\textwidth]{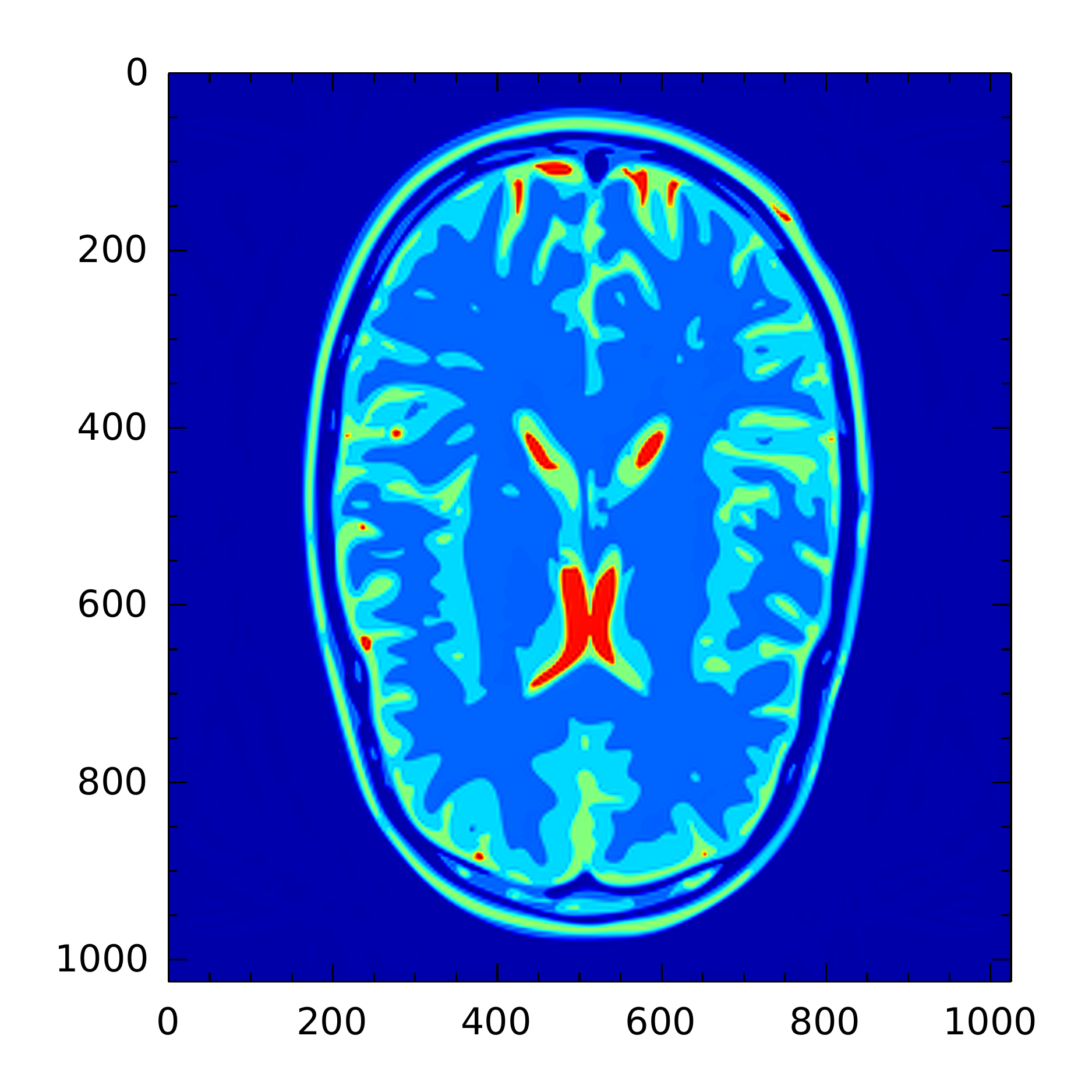}
	\caption{Representation in $256\times 256$ Daubechies 4 scaling functions from $512\times 512$ frequency measurements on a uniform grid. The representation is evaluated at scale 10, i.e., in $1024^2$ points.}
	\label{fig:DB4_brain}
\end{figure}

As an example we consider reconstruction of a simulated brain made with the Matlab \cite{MATLAB} software released along with \cite{Guerquin-Kern:Lejeune:Pruessmann:Unser:2012} (available at \url{http://bigwww.epfl.ch/algorithms/mriphantom}).
The reconstructed brain with the Daubechies 4 scaling functions is seen in Fig.\nobreakspace \ref {fig:DB4_brain}.

These examples are released along with the code.
To avoid having to compute the frequencies for the brain images we have saved these in a native Julia format.
However, that these files are quite large and \emph{not} release with the source code -- instead they are available from one of the author's website:
\url{http://people.math.aau.dk/~robert/software}.

\subsection{Runtime and technical comparison}

Julia is an interpreted language with a fast JIT compiler.
A consequence is that a function is compiled the first time it is called, causing an overhead in terms of time and memory.
All subsequent calls are, however, without this compiling overhead.
The runtimes reported in this section are \emph{not} for the first run.

The examples were carried out on a normal laptop (2.60 GHz Intel Core i7, 8 GB RAM) running GNU/Linux and are summarized in Table\nobreakspace \ref {tab:Julia_vs_Matlab}.
The background for the experiments are as follows:
We compare the GeneralizedSampling package with the Matlab code released with \cite{Gataric:Poon:2016} (available at \url{http://www.damtp.cam.ac.uk/research/afha/code}).
To the best of our knowledge this is the only other publicly available software for generalized sampling.
In this connection two comparisons are relevant:
Computing the representation of the change of matrix (initialization/``init'') and using this representation to compute the least squares solution (solution/``sol'').
In both cases the initialization step is fast for Haar scaling functions: 
All Fourier transforms have simple, closed-form expressions that are easily vectorized.
For higher order Daubechies scaling functions all computations rely on iterative prodcedures.

In both packages the solution step is based on a conjugate gradient like algorithm, where the computational cost is dominated by multiplication with the change of basis matrix and its adjoint.
In Matlab the built-in \lstinline{lsqr} function is used and in Julia a custom implementation of the conjugate gradient algorithm is used.

\begin{table}
	\centering
	\begin{tabular}{r >{$}c<{$} lcrrrc}
		\toprule
		Problem       & \multicolumn{1}{c}{Size} & Language & init ($s$) & sol ($s$) & Iter. ($n$) & $s/n$
		\\
		\midrule
		Uniform 1D    & 8192\times 4096          & Matlab   & 15.0       & 0.13      & 9           & 0.01
		\\
		              &                          & Julia    & 0.34       & 0.04      & 12          & 0.003
		\\
		\midrule
		Jitter 1D     & 5463\times 2048          & Matlab   & 10.0       & 0.28      & 20          & 0.13
		\\
		              &                          & Julia    & 0.23       & 0.08      & 20          & 0.004
		\\
		\midrule
		Uniform 2D    & 512^2\times 256^2        & Matlab   & 0.96       & 5.2       & 9           & 0.58
		\\
                      &                          & Julia    & 0.15       & 17.6      & 16          & 1.10
		\\
		\midrule
		Jitter 2D     & 26244\times 32^2         & Matlab   & 104.8      & 8.3       & 50          & 0.17
		\\
		              &                          & Julia    & 2.4        & 2.4       & 18          & 0.13
		\\
		\midrule
		Spiral        & 27681\times 32^2         & Matlab   & 107.1      & 3.7       & 17          & 0.21
		\\
		              &                          & Julia    & 2.8        & 2.2       & 16          & 0.14
		\\
		\bottomrule
		
	\end{tabular}
	\caption{Runtime comparisons with the Matlab implementation from \cite{Gataric:Poon:2016}. ``Size'' refers to the change of basis matrix and ``Iter.'' is the number of iterations by the iterative solver. In all cases the Daubechies 4 scaling functions are used.}
	\label{tab:Julia_vs_Matlab}
\end{table}

In GeneralizedSampling we have relied on Julia's ability to write functions that modify their arguments in-place to drastically reduce the memory consumption in an iterative algorihtm like conjugate gradients.
Julia's \lstinline{@time} macro makes it easy to estimate the memory allocation of a function.
Matlab has no such documented features and we have therefore not included comparisons on memory usage.

Especially for fast runtimes it is not accurate to rely on timing a single run of a function (using e.g.\ \lstinline{tic} and \lstinline{toc} in Matlab).
In Matlab the times are obtained with the built-in \lstinline{timeit} function and in Julia we use the benchmark package \lstinline{BenchmarkTools} available at \url{https://github.com/JuliaCI/BenchmarkTools.jl}.

For small problems the Julia and Matlab code are comparable, but for large problems the Julia code is significantly faster.
The one execption is for the ``Uniform 2D'' example: 
The Julia timing is using the general NFFT algorithm, whereas the Matlab timing is considering the special case where the standard FFT is applicable, as explained in Section\nobreakspace \ref {sec:NFFT}.

Note that the ``sol'' times and number of iterations are not directly comparable, since \cite{Gataric:Poon:2016} use $L^2([0,1]^d)$ as the reconstruction space and the stopping criteria for the least squares solver may be different.
But the time per iteration (``$s/n$'') are comparable.

\subsection{Availability of the package}

The GeneralizedSampling package is open-source with an MIT license and available from the GitHub repository \url{https://github.com/robertdj/GeneralizedSampling.jl}.
For an easy installation use the built-in package manager in Julia:
\begin{lstlisting}
julia> Pkg.add("GeneralizedSampling")
\end{lstlisting}
This also installs the necessary dependents.

\appendix

\section{Fourier transform of Daubechies wavelets}
\label{sec:Fourier_Daub_wavelets}

In the code and experiments we let $\Hil = L^2(\R)$ and define the Fourier transform of $f \in L^1(\R)$ as
\begin{equation}
	\label{eq:def_Fourier_transform}
	\F[f](k) = \int_\R \exp(-2\pi ikx) f(x) dx.
\end{equation}
Let $\delta_a$ and $\tau_h$ denote a dilation and translation operator, respectively:
$(\delta_a f)(x) = f(ax)$ and $(\tau_h f)(x) = f(x - h)$.
We have the following wellknown relations between the Fourier operator and the dilation and translation operators:
\begin{gather}
	\label{eq:Fourier_dilation}
	\F[\delta_a f](\xi) = \frac1a \F[f]\Bigl(\frac 1a \xi\Bigr), 
	\\
	\label{eq:Fourier_translation}
	\F[\tau_h f](\xi) = \exp(-2\pi i \xi h) \F[f](\xi).
\end{gather}

With the properties \textup {(\ref {eq:Fourier_dilation})} and\nobreakspace  \textup {(\ref {eq:Fourier_translation})} we have for a general scaling function that
\begin{equation}
	\label{eq:Fourier_transform_of_general_wavelet}
	\F[\phi_{j,k}](\xi) 
	= 2^{j/2} \F[\delta_{2^j} \tau_k \phi](\xi)
	= 2^{-j/2} \exp\bigl(-2\pi i k 2^{-j} \xi\bigr) \F[\phi]\bigr(2^{-j} \xi\bigr).
\end{equation}

For the Haar wavelet basis we have closed-form expressions for the Fourier transform of the scaling function:
\begin{gather*}
	\phi(x) = \ind{[0,1)}(x), 
	\\
	\F[\phi](\xi) = 
	\begin{dcases}
		\frac{1 - \exp(-2\pi i\xi)}{2\pi i\xi}, & \xi \neq 0, 
		\\
		1, & \xi = 0.
	\end{dcases}
\end{gather*}

A general Daubechies scaling function $\phi$ is defined by a filter $\Set{h_k}_{k\in\Z}$ where only finitely many entries are non-zero.
The associated low-pass filter, $m_0$, is defined as
\begin{equation*}
	m_0(\xi)
	= \sum_{k\in\Z} h_k \exp(-2\pi i k \xi)
\end{equation*}
The Fourier transform is computed in terms of the low-pass filter:
\begin{equation}
	\label{eq:Daubechies_Fourier_scaling_function}
	\F[\phi](\xi) 
	= \prod_{j=0}^\infty m_0(2^{-j} \xi),
\end{equation}
see e.g.\ \cite{Hernandez:Weiss:1996}.
To ensure convergence of the product in \textup {(\ref {eq:Daubechies_Fourier_scaling_function})}, the filter coefficients must be scaled such that $m_0(0) = 1$.
In the GeneralizedSampling package we use the filters provided in \cite{Cohen:Daubechies:Vial:1993}.

\subsection{Fourier transform of boundary scaling functions}
\label{sec:Fourier_trans_boundary_scaling_function}

Computation of the Fourier transform of the boundary wavelets of \cite{Cohen:Daubechies:Vial:1993} is described in \cite{Gataric:Poon:2016} and repeated here for completion.

The left boundary scaling functions satisfies the following dilation equation:
\begin{equation}
	\label{eq:dilation_eq_lphi}
	\frac1{\sqrt2} \lphi_k(x)
	= \sum_{l=0}^{p-1} \lH_{k,l} \lphi_l(2x) + \sum_{m=p}^{p+2k} \lh_{k,m} \phi(2x-m),
	\Space
	0\leq k< p
\end{equation}
Applying the Fourier transform to this equation yields that
\begin{equation}
	\label{eq:Fourier_relation_lphi}
	\sqrt2 \F\bigl[\lphi_k\bigr](\xi)
	= \sum_{l=0}^{p-1} \lH_{k,l} \F\bigl[\lphi_l\bigr]\Bigl(\frac\xi2\Bigr) + \widehat\phi\Bigl(\frac\xi2\Bigr) \sum_{m=p}^{p+2k} \lh_{k,m} \exp\bigl(-2\pi i m\xi/2\bigr).
\end{equation}
These equations are collected in vector form by introducing the matrices
\begin{equation*}
	\lU = 
	\frac1{\sqrt2} 
	\begin{bmatrix}
		\lH_{0,0}   & \cdots & \lH_{0,p-1}
		\\
		\vdots      & \ddots & \vdots
		\\
		\lH_{p-1,0} & \cdots & \lH_{p-1,p-1}
	\end{bmatrix},
	\Space
	\lV = 
	\frac1{\sqrt2} 
	\begin{bmatrix}
		\lh_{0,p} & 0 & 0 & 0 & \cdots & 0
		\\
		\lh_{1,p} & \lh_{1,p+1} & \lh_{1,p+2} & 0 & \cdots & 0
		\\
		\vdots
		\\
		\lh_{p-1,p} & \lh_{p-1,p+1} & \multicolumn{3}{c}{\cdots} & \lh_{p-1,3p-2}
	\end{bmatrix}
\end{equation*}
and the vectors
\begin{equation*}
	\lv_1(\xi) 
	= \Bigl[ \F\bigl[\lphi_k\bigr](\xi) \Bigr]_{k=0}^{p-1},
	\Space
	\lv_2(\xi)
	= \Bigl[ \widehat\phi(\xi) \exp(-2\pi i m \xi) \Bigr]_{m=p}^{3p-2}.
\end{equation*}
With this notation \textup {(\ref {eq:Fourier_relation_lphi})} can be written as
\begin{align*}
	\lv_1(\xi)
	& = \lU \lv_1\Bigl(\frac\xi2\Bigr) + V \lv_2\Bigl(\frac\xi2\Bigr)
	\\
	& = \lU^j \lv_1\Bigl(\frac\xi{2^j}\Bigr) + \sum_{\ell=0}^{j-1} \lU^\ell V \lv_2\Bigl(\frac\xi{2^{\ell+1}}\Bigr)
\end{align*}
for all $j\geq 1$.
For $\xi = 0$ this equation simplifies to
\begin{equation*}
	\lv_1(0)
	 = \lU \lv_1(0) + V \lv_2(0).
\end{equation*}
that can be solved with respect to $\lv_1(0)$.
Since $\lv_1$ is continuous, we have for large $j$ that
\begin{equation*}
	\lv_1(\xi)
	\approx \lU^j \lv_1(0) + \sum_{\ell=0}^{j-1} \lU^\ell V \lv_2\Bigl(\frac\xi{2^{\ell+1}}\Bigr).
\end{equation*}

The counterpart of \textup {(\ref {eq:dilation_eq_lphi})} for the right boundary scaling functions are
\begin{equation}
	\frac1{\sqrt2} \rphi_k(x)
	= \sum_{l=0}^{p-1} \rH_{k,l} \rphi_l(2x) + \sum_{m=p}^{p+2k} \rh_{k,m} \phi(2x+m+1).
\end{equation}
Introduce the matrices $\rU$ and $\rV$ completely analogously to $\lU$ and $\lV$, respectively, and let
\begin{equation*}
	\rv_1(\xi) 
	= \Bigl[ \F\bigl[\rphi_k\bigr](\xi) \Bigr]_{k=0}^{p-1},
	\Space
	\rv_2(\xi)
	= \Bigl[ \widehat\phi(\xi) \exp\bigl(2\pi i (m+1) \xi\bigr) \Bigr]_{m=p}^{3p-2}.
\end{equation*}
With these notational counterparts, the computations above for the left boundary scaling functions can be copied for the right scaling functions.

\subsection{Fourier transform in 2D}

When considering two dimensional wavelets we introduce the scaling function and the horizontal, vertical and diagonal wavelets as the tensor products
\begin{equation*}
	\phi(\vv x) 
	= \phi(x_1) \phi(x_2),
	\Space
	\psi^1(\vv x) 
	= \phi(x_1) \psi(x_2),
	\Space
	\psi^2(\vv x) 
	= \psi(x_1) \phi(x_2), 
	\Space
	\psi^3(\vv x) 
	= \psi(x_1) \psi(x_2).
\end{equation*}
We denote the two dimensional wavelet functions by
\begin{equation*}
	\phi_{j,\vv n}(\vv x) 
	= 2^j \phi(2^j x_1 - n_1, 2^j x_2 - n_2).
\end{equation*}
When the scale is fixed, the translations are used to index the function.
The separable nature of these functions gives the identity
\begin{equation}
	\label{eq:Fourier_transform_of_2D_wavelet}
	\F[\phi](\vv\xi) 
	= \F[\phi](\xi_x) \F[\phi](\xi_y).
\end{equation}

\section{Weights in non-uniform sampling}
\label{sec:2D_weights_density}

The bandwidth area is divided into the Voronoi tesselation induced by the sampling points, i.e., the Voronoi cell of point $\vv\xi_i$ is
\begin{equation*}
	V_i = \Set[\big]{\vv x \in Y_K \given \abs{\vv\xi_i - \vv x} < \abs{\vv\xi_j - \vv x}, i\neq j}
\end{equation*}
Then $V_i \cap V_j = \emptyset$ if $i\neq j$ and
\begin{equation}
	\label{eq:Voronoi_tesselation}
	Y_K = \bigcup_{m=1}^M \overline{V_m}.
\end{equation}
The weight $\mu_m$ of sampling point $\vv\xi_m$ is then the area of $V_m$.
Let $B$ denote the collection of boundaries of the Voronoi cells:
\begin{equation*}
	B = Y_K \setminus \bigcup_{m=1}^M V_m.
\end{equation*}

The density $\delta$ is defined as
\begin{equation*}
	\delta
	= \sup_{\vv x\in Y_K} \inf_{1\leq m\leq M} \norm{\vv\xi_m - \vv x}.
\end{equation*}
Since $Y_K$ is closed, $\delta$ is attained.
Due to \textup {(\ref {eq:Voronoi_tesselation})}, each $\vv x \in Y_K$ lies either in a unique Voronoi cell or in $B$ and $\delta$ is attained at a point in $B$, which is also a closed set.
More precisely, as $B$ is a union of straight line segments, the supremum is attained at one of the corners in $B$.

\section{Multiplication in 2D}
\label{2d_mult}

In 2D we have the block structure for the reconstruction coefficients as considered in  \textup {(\ref {eq:2Dwavelet_border_division})}. 
As an example, consider the hypothetical situation with 2 left, 2 internal and 2 right scaling functions. The matrix for a single frequency $\vv\xi = (\xi_x, \xi_y)$ is then as follows:

\begin{small}
  \begin{equation*}
    \begin{bmatrix}
      \widehat{\lphi_0}(\xi_x) \widehat{\lphi_0}(\xi_y) &
      \widehat{\lphi_0}(\xi_x) \widehat{\lphi_1}(\xi_y) &
      \widehat{\lphi_0}(\xi_x) \widehat{\phi_2}(\xi_y) &
      \widehat{\lphi_0}(\xi_x) \widehat{\phi_3}(\xi_y) &
      \widehat{\lphi_0}(\xi_x) \widehat{\rphi_1}(\xi_y) &
      \widehat{\lphi_0}(\xi_x) \widehat{\rphi_0}(\xi_y)
      \\
      \widehat{\lphi_1}(\xi_x) \widehat{\lphi_0}(\xi_y) &
      \widehat{\lphi_1}(\xi_x) \widehat{\lphi_1}(\xi_y) &
      \widehat{\lphi_1}(\xi_x) \widehat{\phi_2}(\xi_y) &
      \widehat{\lphi_1}(\xi_x) \widehat{\phi_3}(\xi_y) &
      \widehat{\lphi_1}(\xi_x) \widehat{\rphi_1}(\xi_y) &
      \widehat{\lphi_1}(\xi_x) \widehat{\rphi_0}(\xi_y)
      \\
      \widehat{\phi_2}(\xi_x) \widehat{\lphi_0}(\xi_y) &
      \widehat{\phi_2}(\xi_x) \widehat{\lphi_1}(\xi_y) &
      \widehat{\phi_2}(\xi_x) \widehat{\phi_2}(\xi_y) &
      \widehat{\phi_2}(\xi_x) \widehat{\phi_3}(\xi_y) &
      \widehat{\phi_2}(\xi_x) \rphi_1(\xi_y) & \widehat{\phi_2}(\xi_x)
      \widehat{\rphi_0}(\xi_y)
      \\
      \widehat{\phi_3}(\xi_x) \widehat{\lphi_0}(\xi_y) &
      \widehat{\phi_3}(\xi_x) \widehat{\lphi_1}(\xi_y) &
      \widehat{\phi_3}(\xi_x) \widehat{\phi_2}(\xi_y) &
      \widehat{\phi_3}(\xi_x) \widehat{\phi_3}(\xi_y) &
      \widehat{\phi_3}(\xi_x) \widehat{\rphi_1}(\xi_y) &
      \widehat{\phi_3}(\xi_x) \widehat{\rphi_0}(\xi_y)
      \\
      \rphi_1(\xi_x) \widehat{\lphi_0}(\xi_y) &
      \widehat{\rphi_1}(\xi_x) \widehat{\lphi_1}(\xi_y) &
      \widehat{\rphi_1}(\xi_x) \widehat{\phi_2}(\xi_y) &
      \widehat{\rphi_1}(\xi_x) \widehat{\phi_3}(\xi_y) &
      \widehat{\rphi_1}(\xi_x) \widehat{\rphi_1}(\xi_y) &
      \widehat{\rphi_1}(\xi_x) \widehat{\rphi_0}(\xi_y)
      \\
      \widehat{\rphi_0}(\xi_x) \widehat{\lphi_0}(\xi_y) &
      \widehat{\rphi_0}(\xi_x) \widehat{\lphi_1}(\xi_y) &
      \widehat{\rphi_0}(\xi_x) \widehat{\phi_2}(\xi_y) &
      \widehat{\rphi_0}(\xi_x) \widehat{\phi_3}(\xi_y) &
      \widehat{\rphi_0}(\xi_x) \widehat{\rphi_1}(\xi_y) &
      \widehat{\rphi_0}(\xi_x) \widehat{\rphi_0}(\xi_y)
    \end{bmatrix}
  \end{equation*}
\end{small}
In the column-major ordering of matrices used in Julia, this orders the scaling functions first by the $y$-coordinate and then by the $x$-coordinate.

With an overloading of notation, let $L = \Set{1,\ldots,p}$, $I = \Set{p+1,\ldots,N-p-1}$ and $R = \Set{N-p,\ldots,N}$ denote the column indices of the left, internal and right functions, respectively.
The vectorize function $\vecto: \R^{M\times N} \to \R^{MN}$ stack columns of a matrix into a vector and the index $(i,j)$ in $\cb$ is computed as $\idx(i,j) = (j-1)M + i$ in $\vecto(\cb)$.
With these notations the product $\vv y = \cb \vecto(X)$ consists of contributions from each of the nine parts in \textup {(\ref {eq:2Dwavelet_border_division})}:
\begin{equation*}
	y_m 
	= \sum_{n=1}^N \cb_{m,n} \vecto(X)_n
    = \sum_{J_1,J_2 \in \Set{L, I, R}} \sum_{(i,j)\in J_1\times J_2} \cb_{m,\idx(i,j)} X_{i,j}
\end{equation*}
This means that we have three different scenarios to consider:
\begin{itemize}
	\item Multiplication with the internal part $\Int_x\Int_y$.
    \item Multiplication with the ``corner'' parts, $\Left_x\Left_y$, $\Left_x\Right_y$, $\Right_x\Left_y$ and $\Right_x\Right_y$.
    \item Multiplication with the ``side'' parts, $\Left_x\Int_y$, $\Int_x\Right_y$, $\Right_x\Int_y$ and $\Int_x\Left_y$.
\end{itemize}

We handle each of the nine parts separately, although all of the ``corner''  parts are treated similarly and all of the ``side'' parts are treated similarly.
As an example, consider the contribution from $\Right_x\Left_y$:
\begin{equation*}
	\sum_{(i,j)\in R\times L} \cb_{m,\idx(i,j)} X_{i,j}
	= \sum_{(i,j)\in R\times L} \widehat{\rphi_i}(\xi_{m,x}) \widehat{\lphi_j}(\xi_{m,y}) X_{i,j}
	= \sum_{j\in L} \widehat{\lphi_j}(\xi_{m,y}) \sum_{i\in L} \widehat{\rphi_i}(\xi_{m,x}) X_{i,j}.
\end{equation*}
The last sum is recognized as the $(m,j)$'th entry in the product $\Right_x X$ and hence we get
\begin{equation*}
	\sum_{j\in L} \bigl[\Left_y \had (\Right_x X)\bigr]_{m,j},
\end{equation*}
where $\had$ denotes the Hadamard product.
In the ``side'' parts the multiplication with $\Int_x$ and $\Int_y$ are approximated with an NFFT.

In the multiplication $\vv z = \cb^\A \vv u$, let $\idx(i,j) = n$:
\begin{equation*}
	z_n
    = \sum_{m=1}^M \bigl(\cb^\A\bigr)_{n,m} u_m
	= \sum_{m=1}^M \overline{\widehat{\iphi_{i-1}}}(\xi_{m_x}) \overline{\widehat{\iphi_{j-1}}}(\xi_{m_y}) v_m
	= \sum_{m=1}^M \overline{\widehat{\iphi_{i-1}}}(\xi_{m_x}) \vv v_j,
\end{equation*}
where $\vv v_j = \bigl[\overline{\widehat{\iphi_{j-1}}}(\xi_{m_y})\bigr]_{m=1}^M \had \vv u$.
As above, the nine different parts are treated individually, and multiplications with $\Int_x$ and $\Int_y$ are approximated with an NFFT.

\bibliographystyle{plain}
\bibliography{literature}

\end{document}